# A simple model for the small-strain behaviour of soils

José Jorge Nader

Department of Structural and Geotechnical Engineering, Polytechnic School, University of São Paulo

05508-900, São Paulo, Brazil, e-mail: jjnader@usp.br

Abstract: The constitutive equation discussed in this note eliminates some defects of linear elasticity in the description of the small-strain behaviour of soils. It is capable of representing volume changes in pure shear and different values of bulk modulus in compression and expansion. The new equation provides a simple description of soil behaviour at small strains in that it produces linear stress-strain relations that can approximate the initial part of experimental stress-strain curves.

1. Introduction

Linear elasticity is not a good model for soils even at the small-strain level. Consider, for instance, identical isotropic soil samples subject to different stress paths in the triaxial cell. Although the beginning of each particular stress-strain curve can be approximated by a tangent line through the origin, experience shows that different pairs of elastic parameters (say, bulk and shear modulus) would be necessary to represent the small-strain response in the various stress paths. The simplest example of this fact occurs in the isotropic compression or expansion of a normally consolidated sample: the bulk modulus for a subsequent virgin loading is smaller than the bulk modulus for unloading. Another shortcoming of the linear elastic model is that it cannot predict volume changes in pure shear (e.g., volume increase in dense sands, volume decrease in loose sands) as well as values of Skempton's *A* parameter different from 1/3 in the undrained compression of a saturated soil (see, e.g., Lambe and Whitman 1979, for experimental results).

This note presents a constitutive model for the small-strain behaviour of soils that, unlike linear elasticity, is able to predict the occurrence of volume changes in pure shear, the existence of different values of bulk modulus in isotropic compression and expansion and different values of Young's modulus and Poisson's ratio in axial compression and axial extension. On the other hand, like linear elasticity, the proposed model yields linear stress-strain relations in straight stress-paths and so provides a relatively simple approximate representation of the behaviour at small strains.

Throughout the text, we use the soil mechanics sign convention for stresses and strains (compressive stresses and strains are positive etc).

## 2. The new model

As a starting point, consider the isotropic linear elastic equation, as used in soil mechanics:

$$\Delta \mathbf{T} = K I_1(\mathbf{E})\mathbf{I} + 2G\mathbf{E}_d. \qquad (1)$$

$\Delta \mathbf{T}$, $\mathbf{E}$ and $\mathbf{I}$ are, respectively, the effective stress increment tensor, the strain tensor and the identity tensor; $\mathbf{E}_d$ is the deviatoric part of $\mathbf{E}$, $I_1(\mathbf{E})$ is the first invariant of $\mathbf{E}$, $K$ is the bulk modulus and $G$ is the shear modulus. Soils are always subject to initial effective stresses, whether in the field or in laboratory. Eq. 1 thus yields the increment $\Delta \mathbf{T}$ that should be added to the initial effective stress tensor to obtain the final effective stress tensor (applications of linear elasticity to soil mechanics problems is a classical subject; see, e.g., Terzaghi 1943; Davis and Selvadurai 1996).

Now the first term on the right-hand side of Eq. 1 will be replaced by a more general function of $\mathbf{E}$ to produce the new model. The resulting equation reads

$$\Delta \mathbf{T} = \alpha(\mathbf{E})\mathbf{I} + 2G\mathbf{E}_d, \qquad (2)$$

where $\alpha(\mathbf{E}) = (K_C + K_E)I_1(\mathbf{E})/2 + (K_C - K_E)|I_1(\mathbf{E})|/2 + \xi G|I_2(\mathbf{E}_d)|^{1/2}$. $I_2(\mathbf{E}_d)$ denotes the second invariant of $\mathbf{E}_d$. The parameters $K_C$ (compression bulk modulus), $K_E$ (expansion bulk modulus) and $G$ (shear modulus) are positive whereas $\xi$ (dilatancy coefficient) may be negative, positive or null. Their physical meaning will become clear in the examples below. Note that, for $I_1(\mathbf{E}) \geq 0$, $\alpha(\mathbf{E}) = K_C I_1(\mathbf{E}) + \xi G|I_2(\mathbf{E}_d)|^{1/2}$ while, for $I_1(\mathbf{E}) \leq 0$, $\alpha(\mathbf{E}) = K_E I_1(\mathbf{E}) + \xi G|I_2(\mathbf{E}_d)|^{1/2}$. Of course, Eq. 2 reduces to Eq. 1 if $K_C = K_E$ and $\xi$=0.

In the next section the behaviour of the proposed equation will be investigated in typical situations (isotropic compression and expansion, pure shear, axial compression and extension, undrained axial compression). For the analysis of cases in which $\Delta \mathbf{T}$ is given and $\mathbf{E}$ must be found, it is convenient to invert the stress-strain relation expressed by Eq. 2. As $\mathbf{E} = I_1(\mathbf{E})\mathbf{I}/3 + \mathbf{E}_d$, it suffices to write $I_1(\mathbf{E})$ and

$\mathbf{E}_d$ in terms of $\Delta\mathbf{T}$. Clearly, $\mathbf{E}_d = \Delta\mathbf{T}_d / 2G$ (in which $\Delta\mathbf{T}_d$ is the deviator stress increment tensor). Further, it can be concluded that, if $I_1(\mathbf{E}) \geq 0$, then $I_1(\mathbf{E}) = \rho(\Delta\mathbf{T})/K_C$ while, if $I_1(\mathbf{E}) \leq 0$, $I_1(\mathbf{E}) = \rho(\Delta\mathbf{T})/K_E$, where $\rho(\Delta\mathbf{T}) = I_1(\Delta\mathbf{T})/3 - \xi |I_2(\Delta\mathbf{T}_d)|^{1/2}/2$. Hence, the following equation holds:

$$\mathbf{E} = \beta(\Delta\mathbf{T})\mathbf{I} + \frac{1}{2G}\Delta\mathbf{T}_d, \qquad (3)$$

in which $\beta(\Delta\mathbf{T}) = (1/6K_C + 1/6K_E)\rho(\Delta\mathbf{T}) + (1/6K_C - 1/6K_E)|\rho(\Delta\mathbf{T})|$. Therefore, for $\rho(\Delta\mathbf{T}) \geq 0$, $\beta(\Delta\mathbf{T}) = \rho(\Delta\mathbf{T})/3K_C$, whereas, for $\rho(\Delta\mathbf{T}) \leq 0$, $\beta(\Delta\mathbf{T}) = \rho(\Delta\mathbf{T})/3K_E$. In any case, the volumetric strain is given by $\varepsilon_v = I_1(\mathbf{E}) = 3\beta(\Delta\mathbf{T})$.

### 3. Model behaviour in simple cases

#### a. Isotropic compression and expansion

The response of Eq. 2 to an isotropic strain ($\mathbf{E} = \varepsilon\mathbf{I}$) is an isotropic stress increment ($\Delta\mathbf{T} = \sigma\mathbf{I}$). The relation between the volumetric strain ($\varepsilon_v = 3\varepsilon$) and $\sigma$ depends on the sign of $\varepsilon$. It is $\sigma = K_C \varepsilon_v$ in case of compression ($\varepsilon > 0$) and $\sigma = K_E \varepsilon_v$ in case of expansion ($\varepsilon < 0$). Thus, if $K_C \neq K_E$, the model predicts a different linear stress-strain relation in each case.

#### b. Pure shear

For a pure shear defined by the stress increment matrix

$$[\Delta\mathbf{T}] = \begin{bmatrix} 0 & 0 & 0 \\ 0 & 0 & \tau \\ 0 & \tau & 0 \end{bmatrix} \qquad (4)$$

($\tau > 0$), Eq. 3 gives the following strain matrix:

$$[\mathbf{E}] = \begin{bmatrix} \varepsilon & 0 & 0 \\ 0 & \varepsilon & \gamma/2 \\ 0 & \gamma/2 & \varepsilon \end{bmatrix}. \qquad (5)$$

Thus the shear strain γ is related to the shear stress τ through $\gamma = \tau/G$, as in linear elasticity. By introducing $\rho(\Delta \mathbf{T}) = -\xi\tau/2$ in the expression for $\beta(\Delta \mathbf{T})$, the relation between the volumetric strain ($\varepsilon_v = 3\varepsilon$) and τ can be obtained. If $\xi < 0$, then $\varepsilon_v = -\xi\tau/2K_C$, a positive number, indicating volume decrease (contraction). On the other hand, if $\xi > 0$, then $\varepsilon_v = -\xi\tau/2K_E$, which is negative, indicating volume increase (expansion). Therefore, unlike linear elasticity, the present model predicts volume changes in pure shear.

### c. Axial compression and extension

Now the only non-zero stress increment component is, say, $\Delta T_{33} = \sigma$ (axial compression: $\sigma > 0$; axial extension: $\sigma < 0$). According to Eq. 3, the corresponding strain matrix is diagonal, with $E_{11} = E_{22}$. In compression $\rho(\Delta \mathbf{T}) = m\sigma$, with $m = 1 - 3\xi/2\sqrt{3}$, whereas in extension $\rho(\Delta \mathbf{T}) = n\sigma$, with $n = 1 + 3\xi/2\sqrt{3}$.

Table 1 gives the obtained expressions for the Young's modulus ($\sigma/E_{33}$) in compression ($E_C$) and in extension ($E_E$) in terms of the basic parameters, $G$, $K_C$, $K_E$ and $\xi$. The corresponding values of the Poisson's ratio ($-E_{22}/E_{33}$) in compression ($\nu_C$) and in extension ($\nu_E$) are given in all cases by $\nu_C = E_C/2G - 1$ and $\nu_E = E_E/2G - 1$. Note that there are two different expressions to compute both $E_C$ and $E_E$, each expression associated to an interval of values of ξ. In addition, Table 1 gives expressions for the volumetric strain and, in the last column, restrictions that must be imposed, if any, on the basic parameters in order that the values of Young's modulus and Poisson's ratio are positive, as expected on physical grounds.

The meaning of the symbols HC, LC, LE, HE appearing in the first column of Table 1 will now be explained. They constitute a classification based on the value of the dilatancy coefficient that helps understand the behaviour of the material defined by Eq. 2 in regard to volume changes.

The symbols C and E stand for two main material classes related to volume changes in pure shear. Class C ($\xi < 0$): materials that contract in pure shear. Class E ($\xi \geq 0$): materials that expand ($\xi > 0$) or suffer no volume change ($\xi = 0$) in pure shear.

Classes C and E are further divided into two subclasses according to the behaviour in axial compression and extension. Subclass HC ($\xi < -2\sqrt{3}/3$; high contractibility): materials in class C that contract both in axial compression and extension. Subclass LC ($-2\sqrt{3}/3 \leq \xi < 0$; low contractibility;): materials in class C that contract in axial compression and expand ($-2\sqrt{3}/3 < \xi$) or suffer no volume change $(-2\sqrt{3}/3 = \xi)$ in axial extension. Subclass LE ($0 \leq \xi \leq 2\sqrt{3}/3$: low expansibility): materials in class E that expand in axial extension and contract ($0 \leq \xi < 2\sqrt{3}/3$) or suffer no volume change ($\xi = 2\sqrt{3}/3$) in axial compression. Subclass HE ($\xi > 2\sqrt{3}/3$; high expansibility): materials in class E that expand both in axial compression and extension.

The appropriate expressions for $E_C$, $E_E$ and $\varepsilon_v$ in each subclass can be obtained directly from Table 1.

Table 1: Young's modulus and volumetric strain in axial compression and extension.

| Axial compression ($\sigma > 0$; $m = 1 - 3\xi/2\sqrt{3}$) | | | |
|---|---|---|---|
| $\xi \leq 2\sqrt{3}/3$ ($m \geq 0$) HC, LC, LE | $E_C = \dfrac{9GK_C}{3K_C + mG}$ (6) | $\varepsilon_v = m\sigma/3K_C$ (7) ($\varepsilon_v \geq 0$) | $E_C > 0$. $\nu_C > 0$, if $\xi > 2\sqrt{3}/3 - \sqrt{3}K_C/G$. |
| $\xi > 2\sqrt{3}/3$ ($m < 0$) HE | $E_C = \dfrac{9GK_E}{3K_E + mG}$ (8) | $\varepsilon_v = m\sigma/3K_E$ (9) ($\varepsilon_v < 0$) | $E_C > 0$ and $\nu_C > 0$, if $\xi < 2\sqrt{3}G/3 + 2\sqrt{3}K_E$. |
| Axial extension ($\sigma < 0$; $n = 1 + 3\xi/2\sqrt{3}$) | | | |
| $\xi < -2\sqrt{3}/3$ ($n < 0$) HC | $E_E = \dfrac{9GK_C}{3K_C + nG}$ (10) | $\varepsilon_v = n\sigma/3K_C$ (11) ($\varepsilon_v > 0$) | $E_E > 0$ and $\nu_E > 0$, if $\xi > -2\sqrt{3}G/3 - 2\sqrt{3}K_C$. |
| $\xi \geq -2\sqrt{3}/3$ ($n \geq 0$) LC, LE, HE | $E_E = \dfrac{9GK_E}{3K_E + nG}$ (12) | $\varepsilon_v = n\sigma/3K_E$ (13) ($\varepsilon_v \leq 0$) | $E_E > 0$. $\nu_E > 0$, if $\xi < -2\sqrt{3}/3 + \sqrt{3}K_E/G$. |

**d. Undrained axial compression**

Here the strain matrix reads:

$$[\mathbf{E}] = \begin{bmatrix} -\varepsilon/2 & 0 & 0 \\ 0 & -\varepsilon/2 & 0 \\ 0 & 0 & \varepsilon \end{bmatrix}, \qquad (14)$$

where $\varepsilon > 0$ is the axial compressive strain. From Eq. 2, the non-zero stress increment components read (recall they are effective stresses):

$$\Delta T_{11} = \Delta T_{22} = -(1 - \frac{\sqrt{3}}{2}\xi)G\varepsilon, \qquad (15)$$

$$\Delta T_{33} = (2 + \frac{\sqrt{3}}{2}\xi)G\varepsilon. \qquad (16)$$

In the traditional undrained compression test the total lateral stress remains constant. Hence the pore pressure increment $\Delta u$ is the negative of the lateral effective stress increment:

$$\Delta u = (1 - \frac{\sqrt{3}}{2}\xi)G\varepsilon. \qquad (17)$$

It is positive if $\xi < 2\sqrt{3}/3$ (HC, LC, LE). Only soils in the class HE generate negative pore pressures. The corresponding Skempton's pore pressure parameter is $A = 1/3 - \sqrt{3}\xi/6$.

## 4. Final remarks

Clearly, advanced constitutive models can correct the mentioned defects of linear elasticity and, in addition, are able to describe soil behaviour in complex stress paths (for fundamentals of elastoplastic models see Davis e Selvadurai 2005; as for hypoplastic models, Kolymbas 2000, Nader 2003, 2010). But, due to its simplicity, the proposed model may be advantageous if we are interested only in modelling small-strain behaviour in monotonic short stress-paths, far from failure. Besides, the model may have a didactic value: it can be an intermediate stage between linear elasticity and more complex theories.